# Observation of Majorana Plasmon by Molecular Topological Superconductor and Its Topological SPASER

*Kyoung Hwan Choi[1], and Dong Hack Suh[1†]*

*1 Advanced Materials & Chemical Engineering Building 311, 222 Wangsimni-ro, Seongdong-Gu, Seoul, Korea, E-mail: dhsuh@hanyang.ac.kr*

**Abstract**

Plasmons, quantized collective oscillations of electrons, have been observed in metals and semiconductors. Such massive electrons have been the basic ingredients of research in plasmonics and optical metamaterials.[1] Also, Dirac plasmons have been observed in graphene, two-dimensional electron systems and topological insulators (TIs). A nontrivial $Z_2$ topology of the bulk valence band leads to the emergence of massless Dirac fermions on the surface in TIs.[2,3] Although Dirac plasmons can be formed through additional grating or patterning, their characteristics promise novel plasmonic metamaterials that are tunable in the terahertz and mid-infrared frequency ranges.[4]

Recently, the Majorana fermions have been verified through various kinds of topological superconductors(TSCs). In particular, the quantized and paired spin waves have been discovered in polyaromatic hydrocarbons(PAHs)[5] and Majorana hinge and corner modes have been identified in the organic crystal of PAHs. Interestingly, regularity and periodicity can serve in the xy-plane of the crystal as the patterning of TSC resonators. Here, first we report experimental evidence of Majorana plasmonic excitations in a molecular topological superconductor (MTSC). It was prepared from MTSC resonators with different stacked numbers of HYLION-12. Distributing carriers into multiple MTSC resonators enhance the plasmonic resonance frequency and magnitude, which is different from the effects in a conventional semiconductor superlattice.[6,7] The direct results of the unique carrier density scaling law of the resonance of massless Majorana fermions is demonstrated. Moreover, topological surface plasmon amplification by stimulated emission of radiation (SPASER) is also firstly created from the MTSC resonator. It has two mutually time-reversed chiral surface plasmon modes carrying the opposite topological charges.

**Introduction**

Plasmonics have emerged as a promising path toward nanoscale light circuitry[8,9] and optical integration,[10,11] providing significant subwavelength confinement[12] and mechanisms for both spectral and spatial control over light propagation.[13,14] Traditional plasmonic materials are nobel metals, which are molded into desired geometries by either chemical or lithographic methods.[15] With the advances in colloidal synthesis and free carrier properties in semiconductors, localized surface plasmon resonance(LSPR) properties have now been actively explored in several nano-crystal material classes; metal oxide, metal chalcogenides, metal nitride, and silicon(Si).[16-19]

On the other hands, SPASER has recently experienced rapid development.[20,21] SPASER are similar to lasers, but with a different mechanism. In other words, light is amplified through surface plasmons for SPASER and through population inversion for lasers.[22-28] Some SPASERs are obtained from plasmonic crystals with gain media.[29-31] These are classified as lasing SPASERs, and nanostructured plasmonic metasurfaces consisted of a periodic lattice of individual SPASERs. Due to interactions in the near field, the individual SPASER lock-in phases generate temporally and spatially coherent fields. Various 2D materials with good luminescence properties have been developed including graphene, transition metal dichalcogenides(TMDs) and black phosphorene(BP).[32-34] Dirac plasmons of graphene, the density waves of Dirac fermions, have been demonstrated to enable low loss and effective wave localization up to mid-IR frequencies.[35,36]

The discovery of TIs [37-39] can be served as a novel platform for Dirac plasmon.[3,40] This feature enables topological superconductors, states of much more exotic matters.[41] These topological states of matters are characterized by topological structures of the quantum mechanical wave functions in the Hilbert space. In TSCs, nontrivial Z or $Z_2$ topologies of the

superconducting(SC) states lead to the appearance of Majorana fermions on the surface.[42] Majorana fermions are peculiar in that the particles are their own antiparticles. Originally, thought to be a mysterious neutrinos, it has been proven in recent studies.[43,44] However, lasmons of topological superconductors, Majorana plasmons, have not yet been observed.

Novel higher-order topological odd-parity superconductor and its photonics based on pyrene derivatives have been reported.[5,45] Majorana hinge and corner modes from stacked pyrene derivatives are demonstrated by a constant conductance plateau and an odd parity zero bias conductance peak, respectively. Further, Pancharatnam-Berry phase merging and wavefront shaping are also verified. The topological properties of PAH identified through HYLION-12 were expressed along its z-axis accumulation. However, its periodicity and regularity in the xy-plane can be also an excellent platform for evaluating the plasmonics of the Majorana fermion. The reason is that the extra momentum for plasmonics was obtained by the surface with a sub-wavelength grating.[2,3,46] Here, the methodology of analysis for Dirac plasmon is applied to the spin-coated orthorhombic phase, which is designated as $\Phi_{o,spin}$. It effectively demonstrates phenomena of Majorana plasmonics and topological SPASER in the visible region utilizing patterned MTSC.

**Results and Discussion**

Plasmons of $\Phi_{o,spin}$ were investigated with the dispersion relation of massless electron.[3,47] The samples of $\Phi_{o,spin}$ with various spin coating rates were prepared and confirmed by XRD. (Fig. 1b) The total amount of $\Phi_{o,spin}$ on the coated surface was evaluated based on peak intensity at 2.3 deg because $\Phi_{o,spin}$ has edge-on orientation. Therefore, the (100) face on the coated surface represents the amount of stacking islands of MTSC. It showed the highest intensity of $1.40 \times 10^6$ when the sample was spin-coated at 1750 rpm; other samples exhibited varying intensities from $1.32 \times 10^6$ to $0.70 \times 10^6$ depending on the spin-coating rates. The samples are numbered from 1 to 6 in ascending order of intensity at 2.3 deg and designated as $\Phi_{o,spin,\#}$, where # is the sample number (Fig. 1c).

A UV/Vis spectrophotometer was used to investigate the optical characteristics of $\Phi_{o,spin,\#}$. The molecular absorption of HYLION-12 between 324nm and 366nm is characterized by transmittance (Fig. 2a). This absorbance stems from molecular excitation of electrons. Excitation for wavelengths greater than $\lambda_{max}$ is not related with molecular excitation.[48] Therefore, the absorption in a wide range from 500nm to 1500nm wavelength occurs due to the structure of $\Phi_{o,spin}$, rather than characteristics exhibited by MTSC itself.(Fig. 2a, inset)

By further investigating this exotic absorption, transmittance was converted to optical extinction. (Fig. 2b) The increased intensity of optical extinction results from the increased number of oscillators, and/or the increasing wave number is caused by dipole friction between HYLIONs-12.[47] Additionally, the shape of the optical extinction was slightly asymmetric. These asymmetric patterns may be due to the anisotropic pyrene moiety of HYLION-12. Plasmons exhibit various plasma frequencies depending on the shape and arrangement of the particles.[49]

Collective excitations in regular structures have typically been studied by extracting meaningful constants from the optical extinction patterns. The optical extinction (1-T/T$_0$) is related to the dynamic conductivity $\sigma(\omega)$, as follows:

$$1 - \frac{T}{T_0} = 1 - \frac{1}{|1+Z_0\sigma(\omega)/(1+n_s)|^2} \quad (1)$$

where $Z_0$ is the vacuum impedance, $\omega$ is the frequency, and $n_s$ is the refractive index of the substrate. Equation (1) was used to analyze the plasmon of patterned graphene sheet in the circle.[50-52] In these proactive studies, the complex optical conductivity based on the Drude model is extracted from the optical variants. Additionally, the dynamic conductivity that neglects the lateral disk-disk interaction in a quasi-static approximation was used to analyze the plasmon of patterned graphene to determine the optical conductivity.[47,53,54] The difference between previous studies and present works of $\Phi_{o,spin,\#}$ occurs in the shape of the building block. Therefore, the shape factor *s* is introduced to represent the dynamic conductivity and is used to analyze asymmetric optical extinction:

$$\sigma(\omega) = i\frac{SfD}{\pi}\frac{\omega}{(\omega^2-\omega_p^2+i\Gamma_p\omega)} \quad (2)$$

where ω is the frequency, $f$ is the filling factor, D is the Drude weight, and $\Gamma_p$ is the plasmon resonance width.

Each optical extinction of $\Phi_{o,spin,\#}$ is fitted with equation (1) (Fig. S1 – S6), and the fitting results clearly demonstrate the effectiveness of introducing the shape factor. The shape factors for all samples were between 1.33 and 1.44, which agrees well with the calculated aspect ratio of a long to short axis of the pyrene derivative, i.e., 1.44. Additionally, the scattering width remains constant regardless of the quantity of $\Phi_{o,spin}$ (Fig. 2c). Moreover, plasmon frequencies of $\Phi_{o,spin,\#}$ increase as the amount of the (100) face increases; it is an inherent tendency of plasmons (Fig. 2d).[47]

This significant upshift of the plasmon frequency is due to the strong Coulomb interaction between the adjacent layers. The collective motion of the carriers among the layers leads to more vital restoring force through the dipole-dipole coupling,[55] and the increasing tendency of the plasma frequency is consistent with that of the amount of $\Phi_o$, clearly demonstrating the typical characteristics of the plasmon.

Additional experiments should be conducted to determine the state of the electrons that produce the plasmons of HYLION-12. The non-trivial state of an electron can be determined by analyzing the correlation between Fermi level and plasmon frequency.[56] The relationship between plasma frequency and the number of electrons shows various characteristics depending on the physical properties of the electrons.[57] Especially, the relationship between normalized Fermi level and normalized plasmon frequency is consistent with the characteristics of the electrons.[3] According to this, the normalized plasma frequency and the normalized Fermi level have a square root relationship in the case of a massless electrons, while a linear relationship in the case of a massive electron.

The Fermi level of each sample can be calculated using the HOMO level from UPS and the band gap from a UV/Vis spectrophotometer.[58] The HOMO level was confirmed by UPS according to the previously reported method, and these results demonstrate the increasing tendency of the HOMO level based on the quantity of the (100) face; the HOMO level of $\Phi_{o,spin,1550}$ starts at 3.8 eV and increases gradually. (Fig. 3a) The plot of normalized Fermi level versus normalized plasmon frequency clearly shows that the electrons of aromatic molecules are massless; this relationship clearly obeys the square root law. (Fig. 3b)

On the other hand, these characteristics shows its possibility of $\Phi_{o,spin,\#}$ as a topological SPASER, because it exhibits the various topological photonic phenomena over a wide wavelength range from UV to visible. A diluted $CHCl_3$ solution of HYLION-12 depicts two strong emission peaks at 390 nm and 403 nm for the excitation wavelength range of 230 to 350 nm. (Fig. 4a) It is a typical excitation-emission pattern of an organic molecule. No excitation peak was observed above $\lambda_{max}$. However, $\Phi_{o,spin,\#}$ shows a very different pattern in its diluted solution. Even if excitation experiments were conducted for various wavelengths from 200nm to 500nm, no emission peak was observed for wavelengths above $\lambda_{max}$. (Fig. 4b)

On contrary, excitation of $\Phi_{o,spin,\#}$ has the long excitation tail in a wavelength longer than $\lambda_{max}$. (Fig. 4c) It comes from both of the MTSC resonator and its structure. Another peculiar phenomenon is that emission of the identical wavelength is strongly conducted upon excitation at 540nm where molecular excitation does not occur. To analyze this phenomenon in more detail, various excitation-emission spectra are obtained. (Fig. 4d) Emission of the same wavelength as excitation is observed in the range of 200nm to 500nm, which cannot be observed in the typical molecular excitation-emission spectra. Moreover, second emission with half the frequency was identified. These unusual results lead to several questions: 1) why does a molecular excitation-emission relationship disappear in HYLION-12 and a new relationship appear in $\Phi_{o,spin,\#}$. 2) what is the reason for auto-tunability and expansion in the emission wavelength range in the new relationship? 3) why is the enhancement of emission so strong? All these questions can be easily answered, if $\Phi_{o,spin}$ is assumed as topological SPASER. The first exotic phenomenon is closely related to the mechanism of SPASER, and the energy transition phenomenon clearly shows that the exciton of an electron-hole pair transfers energy to plasmon resonance.[59] Therefore, the disappearnace of the usual molecule's excitation-emission relationship is a clue that plasmon mode is occurring. And it shows the appearance of various plasmon frequencies is related to the structure of $\Phi_{o,spin}$. It is well known that graphene undergoes changes in plasmon frequency when the polyaromatic hydrocarbons grow in size or form stacks.[60-62] Furthermore, red-shifting and blue-shifting of plasma frequencies occur in longitudinal and transverse directions, respectively. In $\Phi_{o,spin}$, it is natural to have a Lorentzian distribution of HYLION-12 stacked perpendicularly; it leads to various kinds of plasmon frequencies. Besides, emission of HYLION-12 is enhanced by plasmons. The unit-cell size of the xy-plane in $\Phi_o$ is small enough to affect each other like a cavity structure. As a result, stimulated emission is observed. And $\Phi_{o,spin}$ can be thought of as an auto-tunable topological SPASER by Majorana plasmon.

**Conclusion**

In conclusion, we have reported the first observation of Majorana plasmon excitation in $\Phi_{o,spin}$ stacked from HYLION-12 which is a molecular topological superconductor (MTSC). Unlike Dirac plasmon in patterned TI which has to go through additional steps for patterning or grating, $\Phi_{o,spin}$ intrinsically act as an array of MTSC resonators. A direct result of the unique carrier density scaling law of massless Majorana fermions is demonstrated using the relationship between the Fermi level and the plasma frequency. Furthermore, the topological SPASER is first demonstrated in the UV/vis region and it can be output without a corresponding coupler. It shows the potential of the MTSC resonator based on dielectric materials for plasmonic applications. Topological SPASER of MTSC also shows a non-linear property that creates second emission with half the frequency of first emission.

Figures

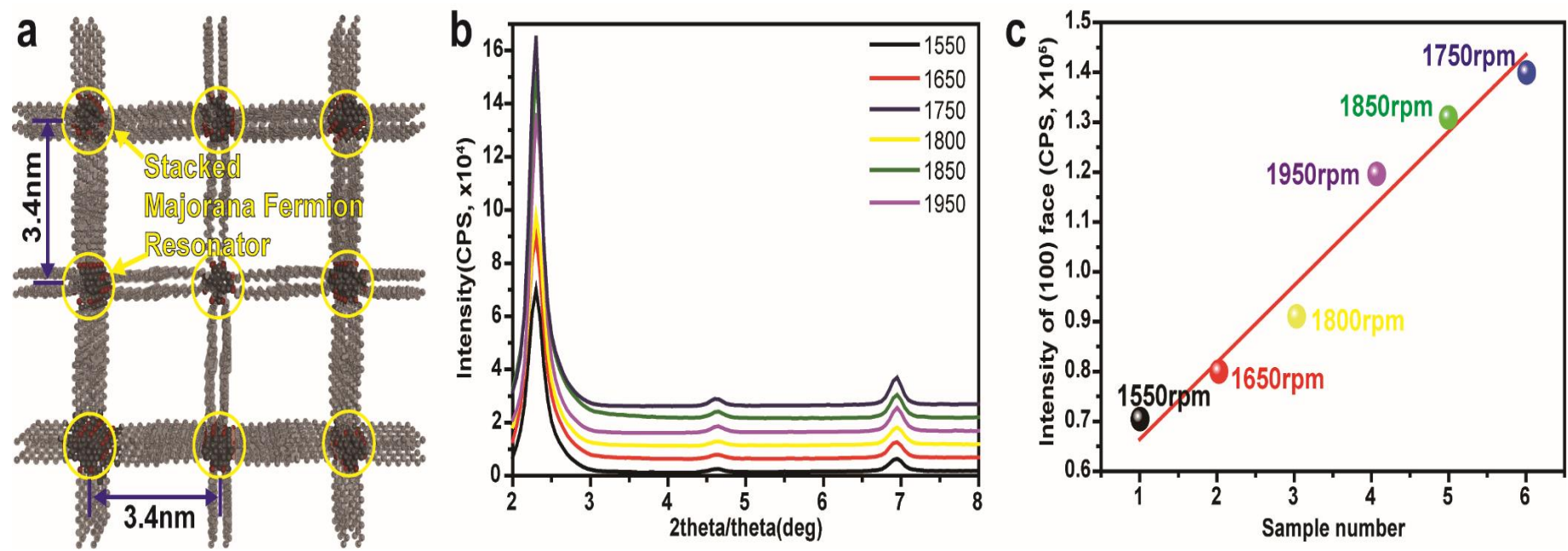

**Figure 1. Structure of Majorana fermion plasmon of orthorhombic phase of HYLION-12. a.** Scheme of Majorana plasmons based on stacked Majorana fermion resonators based on pyrene derivative. b. PXRD patterns of spin-coated HYLION-12 with various spin coating rates: 1550(black), 1650(red), 1750(navy), 1800(yellow), 1850(green), and 1950(magenta) c. Intensities of (100) face of spin-coated orthorhombic phase with various values depending on spin-coating rates: 1550rpm(1, black), 1650rpm(2, red), 1800rpm(3, blue), 1950pm(4, magenta), 1850rpm(5, green) and 1750rpm(6, navy)

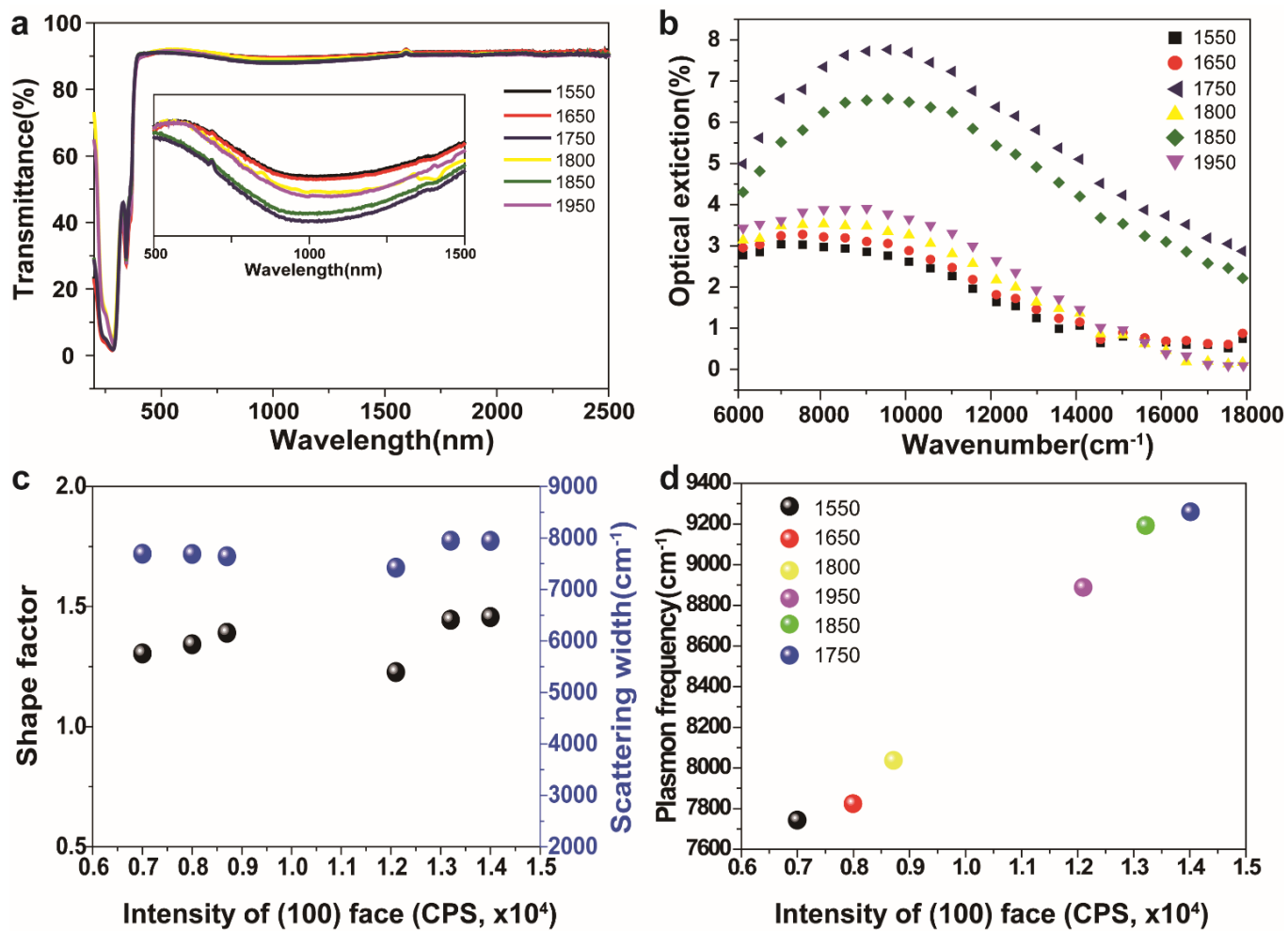

**Figure 2. Optical characteristics and fitting results of $\Phi_{o,spin,\#}$. a.** Transmittance vs. wavelength according to spin-coated rates: 1550(black), 1650(red), 1750(navy), 1800(yellow), 1850(green), and 1950(magenta). **b.** Optical extinction vs. wavelength according to spin-coated rates: 1550(black), 1650(red), 1750(navy), 1800(yellow), 1850(green), and 1950(magenta). **c.** Shape factor and scattering width vs. intensity of (100) face of samples. **d.** Plasmon frequency vs. intensity of (100) face to spin coated rates: 1550(black), 1650(red), 1750(navy), 1800(yellow), 1850(green), and 1950(magenta).

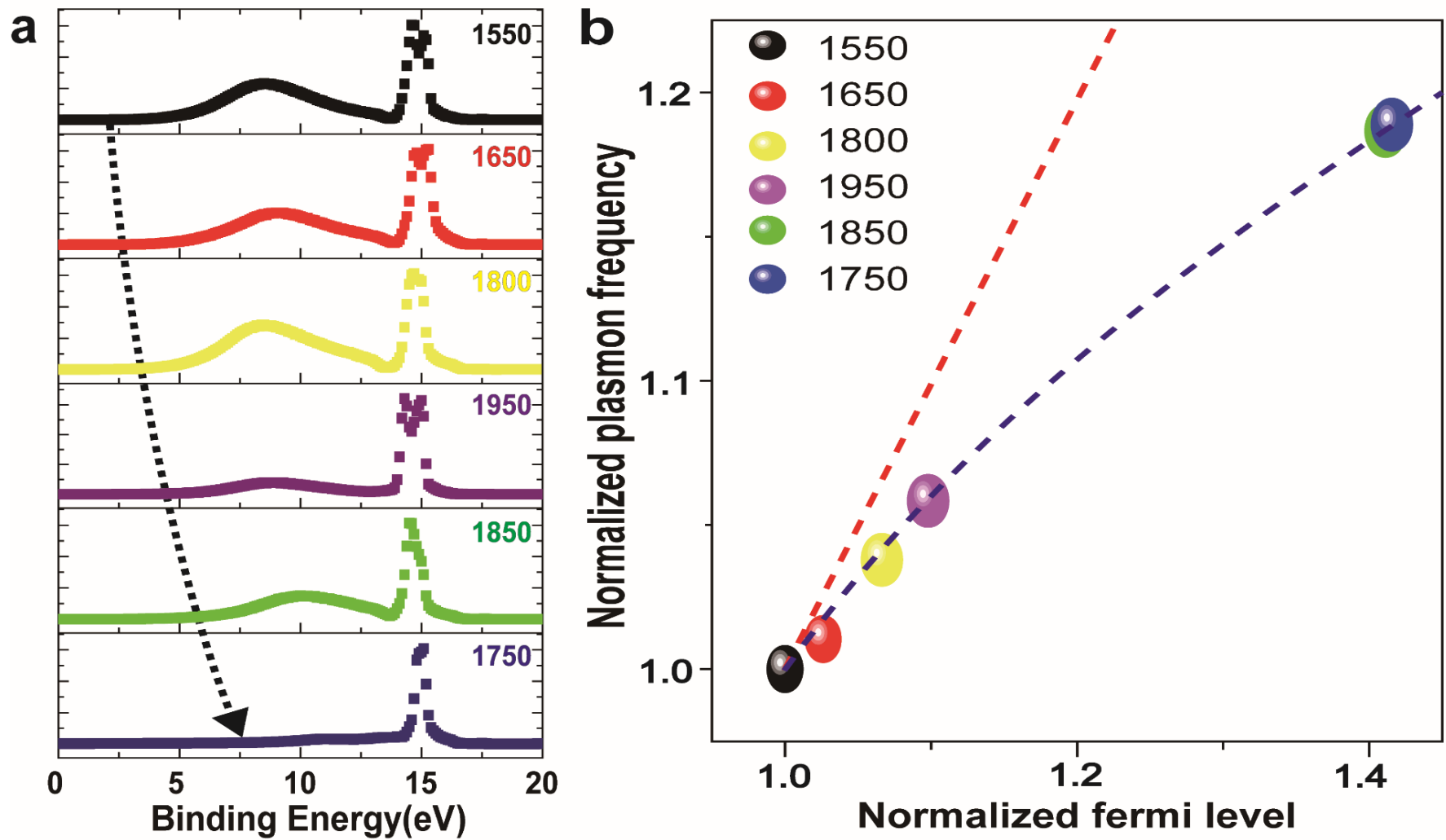

**Figure 3. Majorana plasmon of molecular TSC resonators in $\Phi_o$. a.** UPS analysis of $\Phi_{o,spin,\#}$ according to spin-coated rates: 1550(black), 1650(red), 1750(navy), 1800(yellow), 1850(green), and 1950(magenta). and **b.** Normalized plasmonic resonance frequency as function of normalized Fermi level depicting that direct consequence of unique carrier density scaling law of massless fermions.

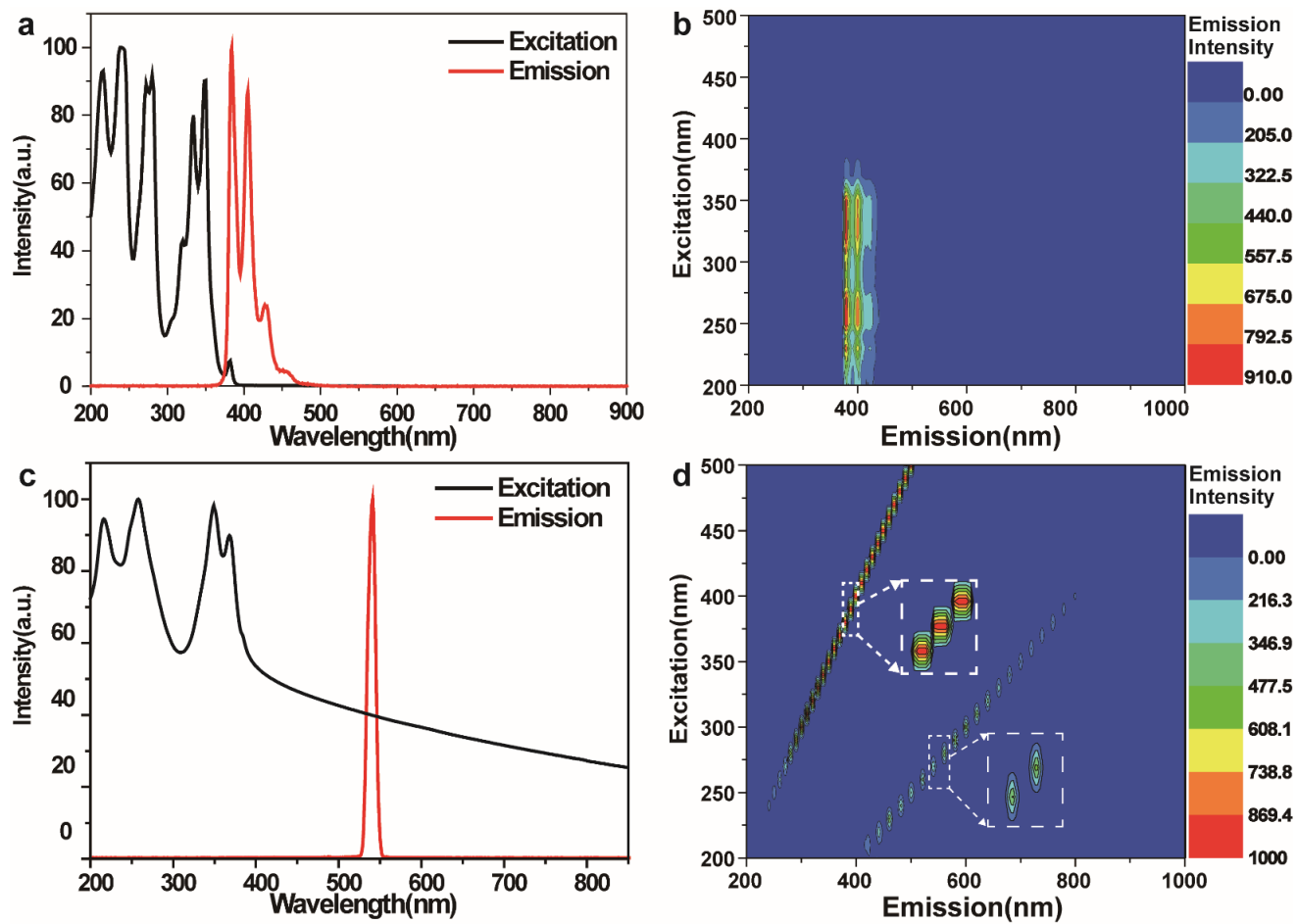

**Figure 4. Photoluminescence analysis. a and b.** Excitation-emission spectra of HYLION-12 solution (0.5 mM, CHCl$_3$) and $\Phi_{o,spin,1}$. The excitation spectra observed from 200 nm to 900 nm for HYLION-12 solution (black line in a) and $\Phi_{o,spin,1}$ (blue line in b). Emission spectrum at 350 nm excitation for HYLION-12 solution (red line in a) and $\Phi_{o,spin,1}$ (black dash-dot in b). **c and d.** Excitation-emission map of HYLION-12 solution (0.5 mM, CHCl$_3$) and $\Phi_{o,spin,1}$. Excitation conducted from 200 nm to 500 nm at 10 nm intervals, and emission measured from 200 nm to 1000 nm.

**Methods**

Pyrene, NaIO4, Bu4NBr, Na2S2O4, and Dodecyl bromide were purchased from Sigma Aldrich and used as received. CDCl3 was purchased from Cambridge Isotope Laboratories and was used for the 1H NMR spectroscopic studies.

A mixture of pyrene-4, 5, 9, 10-tetraone(10mmol), Bu4NBr (13mmol), and Na2S2O4 (115mmol) in H2O (50ml) and THF (50ml) was shaken for 5min. Then the color of mixture was changed from dark brown to pale yellow. Bromododecane (60mmol) was added, and followed by aqueous KOH (306mmol, in 50ml H2O). The mixture was stirred for overnight, poured into a mixture solution of H2O (50ml) and ethyl acetate (30ml). The yellow solid was filtered and washed with ethanol. After drying in a vacuum, the solid was recrystallized from ethyl acetate resulting in a white solid with a yield of 85%.

Spin coating of $\Phi$o. To obtain the orthorhombic phase of HYLION crystal coated onto soda-lime glass, spin-coating was performed with spin coating rate, various solvents and concentrations. Spin coating rate was executed between 1550 rpm and 1950 rpm. A mixture of chloroform and ethyl acetate (2/1v) was selected as the best solvent for the spin-coating process. The concentration of HYLION-12, one of the most important variables for spin-coating, was also adjusted between 10 mM to 50 mM, and a solution of 25 mM was identified as the ideal concentration to construct the orthorhombic phase for each substrate.

The theta/2theta and in-plane X-ray diffraction patterns of all samples were measured on a SMARTLAB (Rigaku Co. Ltd.) diffractometer using monochromatized Cu-K$\alpha$ ($\lambda$ = 0.15418 nm) radiation under 40 kV and 100 mA.

A UV/Vis/NIR spectrometer was used to perform the absorption measurements in transmission geometries. The vertical incidence beam size of the sample was approximately 1 cm$^2$. All measurements were performed at room temperature in nitrogen environments. For graphene on a substrate with an index of refraction ns, the reduction in transmission 1 - $T/T_0$ is related to the complex optical conductivity $\sigma(\omega) = \sigma'(\omega) + i\sigma''(\omega)$ in the following way:

$$1 - T/T_0 = 1 - \frac{1}{|1 + Z_0\sigma(\omega)/(1 + n_s)|^2}$$

where $Z_0$ is the vacuum impedance, $(\mu_0/\varepsilon_0)^{1/2}$, and $\omega$ is the frequency. When the reduction in transmission is small, it only depends on the real part of the conductivity $\sigma'$:

$$\sigma(\omega) = i\frac{SfD}{\pi}\frac{\omega}{(\omega^2 - \omega_p^2 + i\Gamma_p\omega)}$$

The intensity of fluorescence spectra was recorded using Varian-Cary Eclipse fluorescence spectrometer using an excitation wavelength between 200 nm and 500nm wavelength with 10nm step and excitation and emission slit widths of 2.5nm at room temperature.

The Ultraviolet photoelectron spectroscopy (UPS) measurements and Ar gas cluster ion beam (GCIB) sputtering were performed in an ultra-high vacuum analysis system (ULVAC-PHI VersaProbe PHI 5000). A He I (21.2 eV) source was used for UPS measurements, with a −5 V bias to the sample. The GCIB cluster was 2000 Ar atoms/cluster in size, and the accelerating voltage to the single negatively charged cluster was 5 kV. Each UPS spectrum was recorded at the interval of 1 min.

**Acknowledgement**

This research was supported by Basic Science Research Program through the National Research Foundation of Korea(NRF) funded by the Ministry of Education (2018R1D1A1A02047853).

**Author contributions**

Dong Hack Suh designed, initiated and directed this research. Kyoung Hwan Choi performed the experiment and analyzed the data. Dong Hack Suh and Kyoung Hwan Choi co-wrote the manuscript. All authors discussed the results.

**Competing interest.** The authors declare no competing interests.

**Supplementary information** is available for this paper

Supplementary information:

# Observation of Majorana Plasmon by Molecular Topological Superconductor and Its Topological Spaser

*Kyoung Hwan Choi[1], and Dong Hack Suh[1†]*

*1 Advanced Materials & Chemical Engineering Building 311, 222 Wangsimni-ro, Seongdong-Gu, Seoul, Korea, E-mail: dhsuh@hanyang.ac.kr*

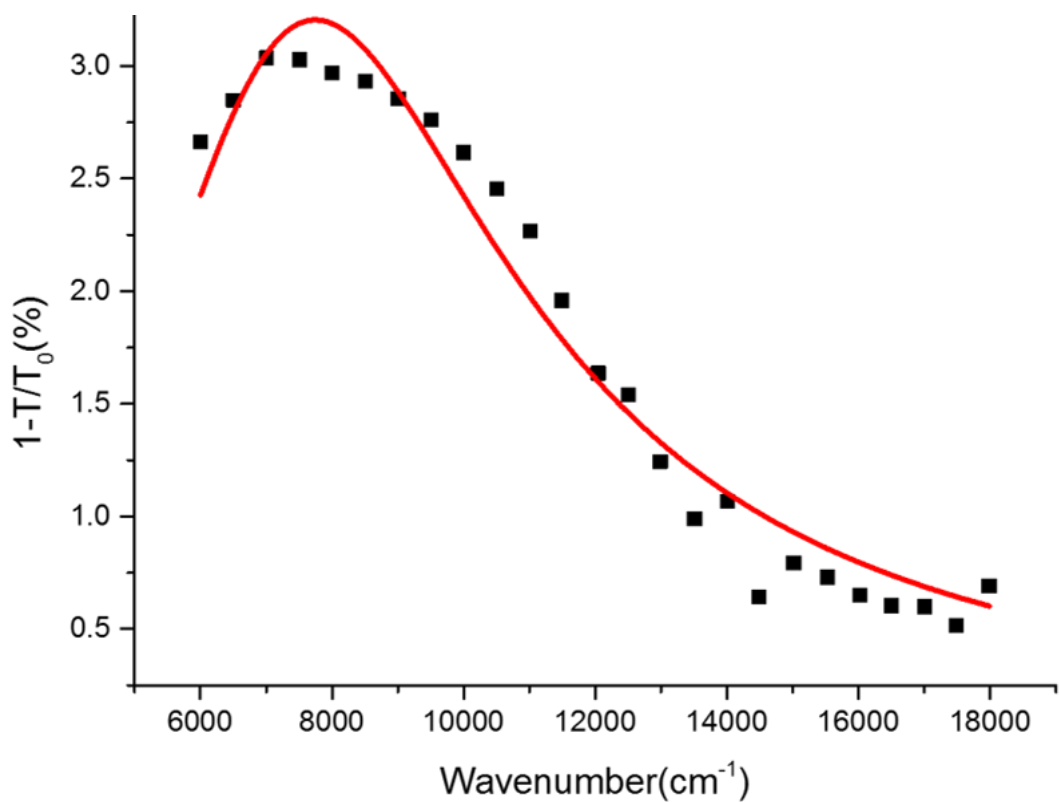

**Figure S1.** Optical extinction of spin coated HYLION-12 on sodalime glass with 1550 rpm and its fitting results.

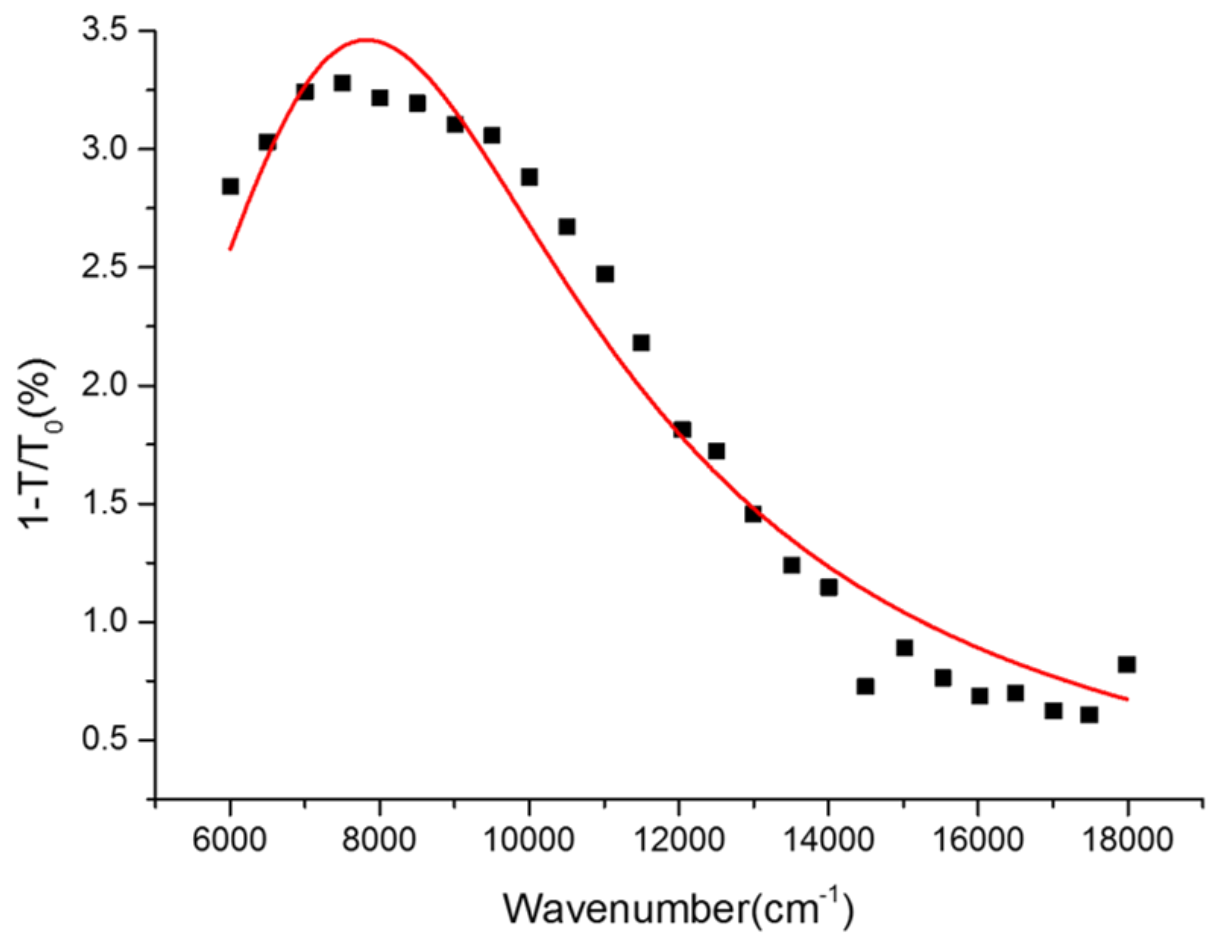

**Figure S2.** Optical extinction of spin coated HYLION-12 on sodalime glass with 1650 rpm and its fitting results.

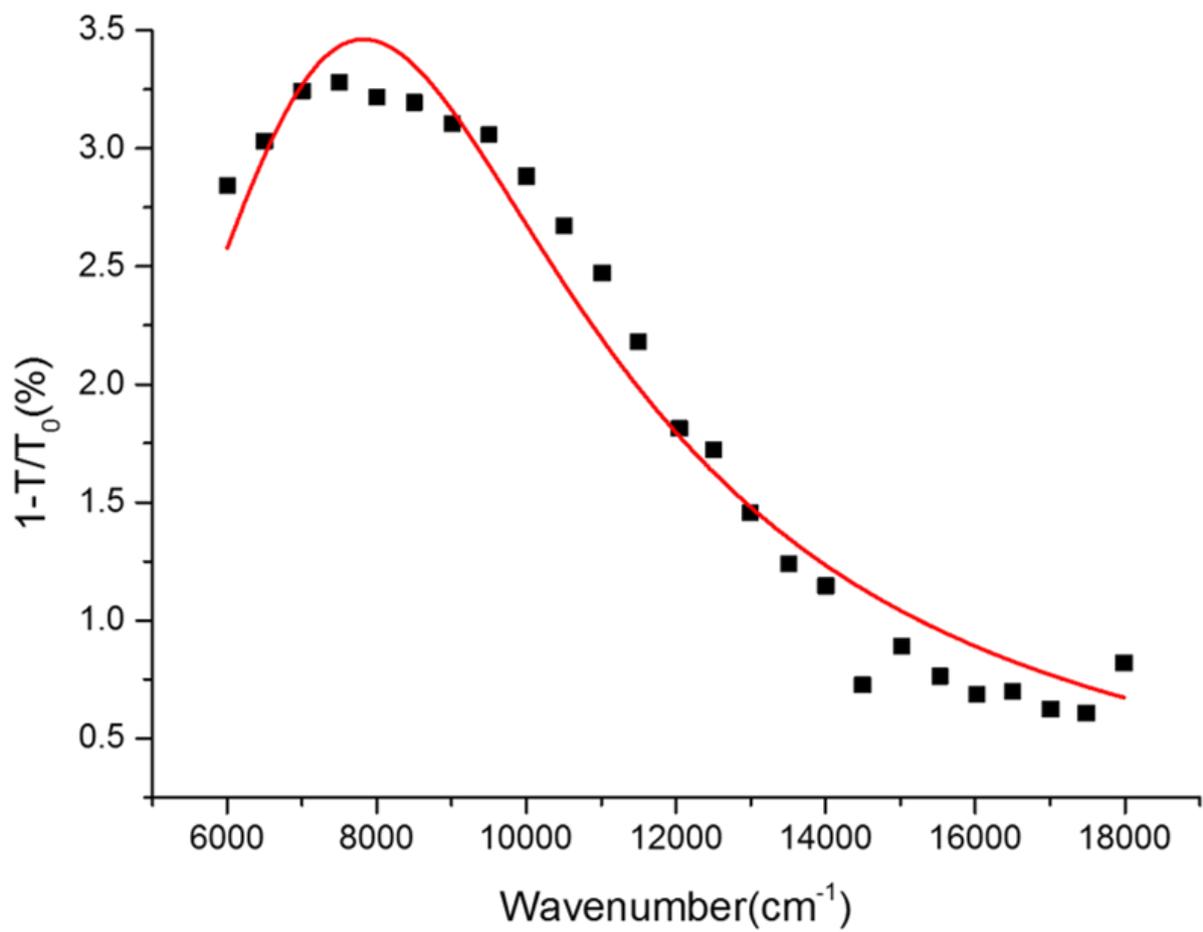

**Figure S3.** Optical extinction of spin coated HYLION-12 on sodalime glass with 1800 rpm and its fitting results.

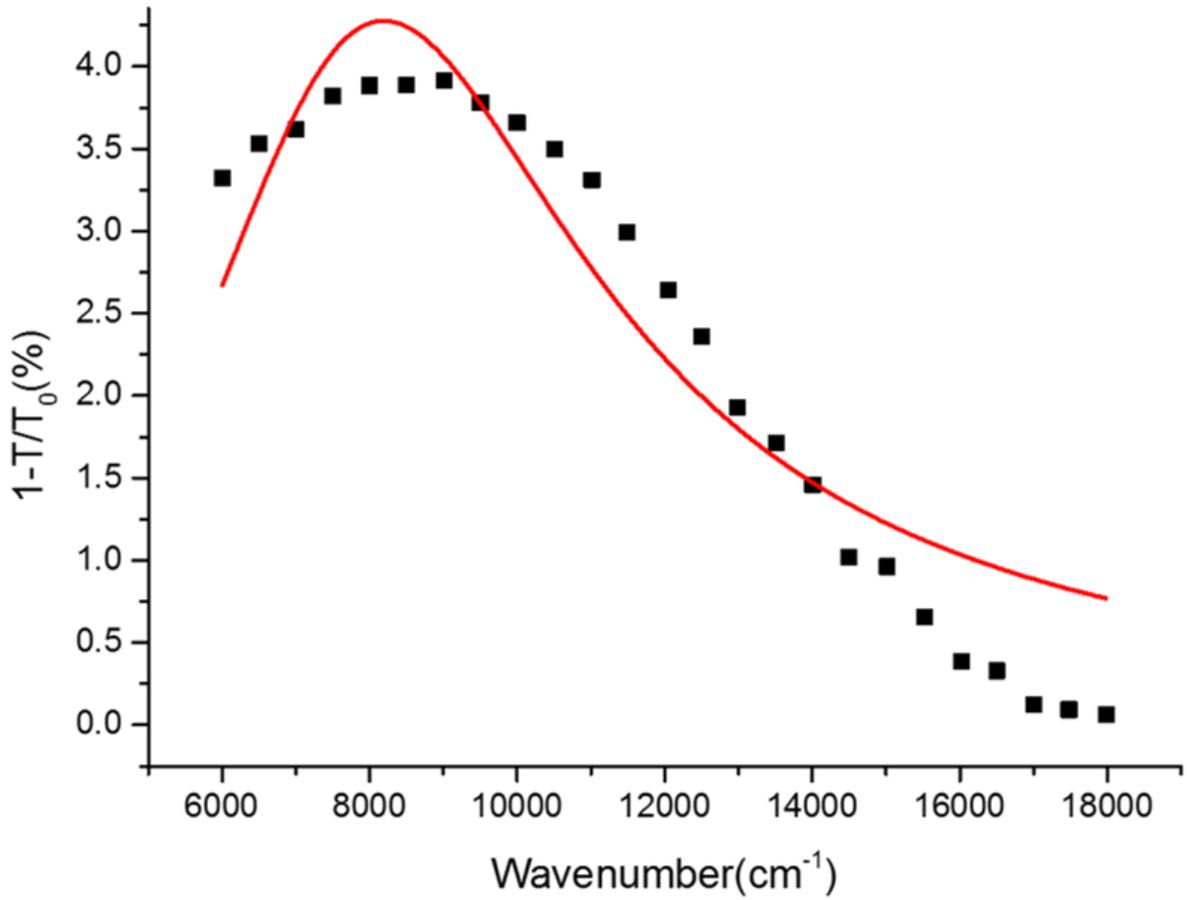

**Figure S4.** Optical extinction of spin coated HYLION-12 on sodalime glass with 1950 rpm and its fitting results.

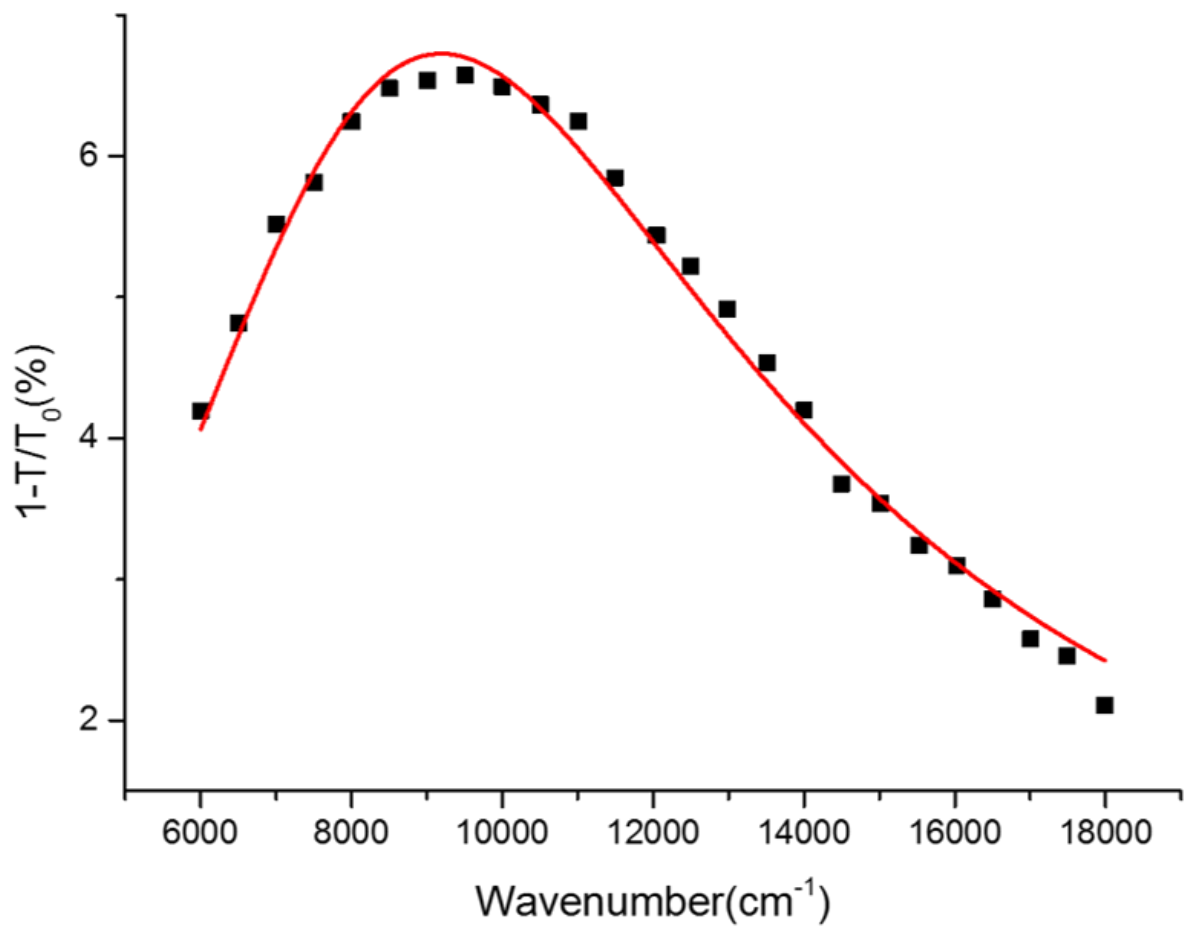

**Figure S5.** Optical extinction of spin coated HYLION-12 on sodalime glass with 1850 rpm and its fitting results. It was designated as sample number 5.

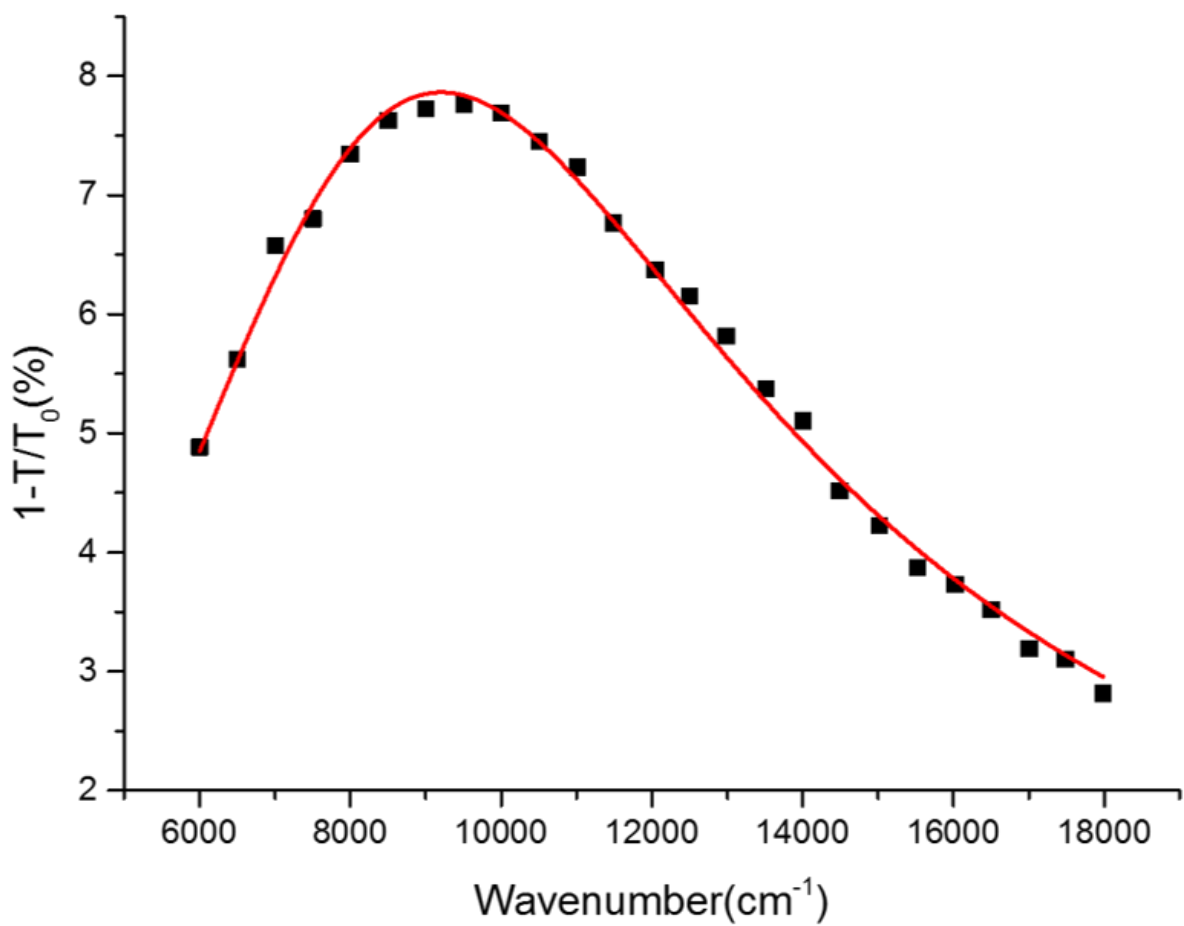

**Figure S6.** Optical extinction of spin coated HYLION-12 on sodalime glass with 1750 rpm and its fitting results.

|   | HOMO | LUMO | Fermi level | Plasmon frequency | Scattering width | Shape factor |
|---|---|---|---|---|---|---|
| **1** | 5.1923 | 1.8047 | 3.4985 | 7743.300 | 6934.5059 | 1.3046 |
| **2** | 5.2522 | 1.8646 | 3.5898 | 7823.789 | 7039.0815 | 1.3430 |
| **3** | 5.3395 | 1.9520 | 3.7297 | 8041.063 | 6467.3309 | 1.3907 |
| **4** | 5.5846 | 2.1971 | 3.8445 | 8192.233 | 6529.9238 | 1.2270 |
| **5** | 6.4995 | 3.1120 | 4.9451 | 9190.647 | 9635.8783 | 1.4449 |
| **6** | 7.4725 | 3.8552 | 4.9512 | 9207.029 | 9889.3467 | 1.4562 |

**Table S1.** Fitting results of UPS and optical extinction.